# Identification of hydrodynamic instability

# by convolutional neural networks

Wuyue Yang[1*], Liangrong Peng[2*], Yi Zhu[1], Liu Hong[1†]

[1]*Zhou Pei-Yuan Center for Applied Mathematics, Tsinghua University, Beijing, 100084, P.R.C.*

[2]*College of Mathematics and Data Science, Minjiang University, Fuzhou, 350121, P.R.C.*

*The co-authors contribute equally to this work.

[†]Email address for correspondence: zcamhl@tsinghua.edu.cn

The onset of hydrodynamic instabilities is of great importance in both industry and daily life, due to the dramatic mechanical and thermodynamic changes for different types of flow motions. In this paper, modern machine learning techniques, especially the convolutional neural networks (CNN), are applied to identify the transition between different flow motions raised by hydrodynamic instability, as well as critical non-dimensionalized parameters for characterizing this transit. CNN not only correctly predicts the critical transition values for both Taylor-Couette (TC) flow and Rayleigh- Bénard (RB) convection under various setups and conditions, but also shows an outstanding performance on robustness and noise-tolerance. In addition, key spatial features used for classifying different flow patterns are revealed by the principal component analysis.

## I. INTRODUCTION

From electron transportation in nano devices to density waves of spiral galaxies, from directional walks of molecular motors driven by ATP hydrolysis to collective motion of bird flocks and fish schools, hydrodynamic flows are the most common yet also the most complicated phenomena in nature. Due to its intrinsic nature about nonlinearity, stochasticity, multiscale, non-homogeneity, etc., a hydrodynamic flow can exhibit completely different behaviors by slightly changing some physical parameters. This dramatic change of flow types is related to the so-called hydrodynamic instability[1], which not only plays an essential role in various branches of fluid mechanics, but also may cause serious consequences. Typical examples are natural disasters, like tsunami and mudslide[2].

Thanks to the pioneering works of Helmholtz, Kelvin, Rayleigh and Reynolds *et al.*, various analytic theories on hydrodynamic stability were developed since the nineteenth century. For example, the Kelvin-Helmholtz instability[3,4], Taylor-Couette instability[5] and Rayleigh-Bénard instability[6] have been analyzed by the linear stability analysis (LSA) with respect to the Navier-Stokes equations. Later, many numerical methods and nonlinear stability theories were developed to analyze the stability of very complicated fluid flows[7,8]. In spite of these progresses, till now the analysis of hydrodynamic stability is still a quite challenging problem in industry, engineering and applied mathematics communities[1,9,10].

It is widely known that the onset of hydrodynamic instability is closely related to the critical values of some dimensionless parameters in the fluid systems, such as the Reynolds number, Rayleigh number, Mach number, *etc.*[1,7]. However, how to determine these critical values is not a trivial question. The frequently used linear stability analysis has a limited accuracy, while the nonlinear stability analysis is usually too complicated to implement. Recently, the explosive development in machine learning algorithms provide an alternative solution, which in principle allows us to identify the place where the hydrodynamic instability takes place, to discriminate laminar flows from turbulence, to predict critical parameters and their influences on hydrodynamic flows. The machine learning algorithms are automatic, efficient, reliable, ready for generalization to new situations, and applicable to highly chaotic systems, like turbulence, in principle.

Previously, many scholars have applied machine learning to discover new phases and phase transitions in condensed matter physics and material science. For example, Carrasquill and Melko[11] showed fully connected and convolutional neural networks can identify highly non-trivial states with no conventional order parameters in a variety of condensed-matter Hamiltonians, by directly sampling raw state configurations of Ising models with Monte Carlo simulations. Similar results were reported for unsupervised learning in Ising model[12] and frustrated XY model[13] too. Unsupervised machine learning method can also be applied to identify the critical state in Ferroelectrics at the nanometer scale[14], though the order parameter is unknown. Note that, the deep neural networks (DNNs) were used to reveal Reynolds averaged Navier Stokes turbulence models[31,32]. Meanwhile, in the field of fluid mechanics, there is no or quite few such works to identify the critical state till now.

In this paper, to illustrate the great potential of machine learning algorithms in hydrodynamic studies, we consider two classical hydrodynamic flows -- Taylor-Couette flow and Rayleigh-Bénard convection. Both of them show typical phenomena of hydrodynamic instabilities with respect to changes of the Reynolds number ($Re$) and Rayleigh number ($Ra$). High-resolution flow fields under diverse $Re$ and $Ra$ values are numerically simulated by mature CFD softwares (OpenFOAM to be exact). Critical $Re$ and $Ra$ numbers for the transition between laminar flows and vortical flows for TC and RB flows are predicted by Convolutional Neutral Networks (CNN) and compared with results based on linear stability analysis. Further data masking and random perturbation experiments show CNN predictions are quite robust and noise-tolerant, which is essential for real applications. Finally, key spatial features are revealed by PCA, which provides further detailed insights on the working mechanism of CNN.

## II. RESULTS and DISCUSSION
### A. Numerical simulations of TC and RB flows

The Taylor-Couette flow is a classic example of a simple system with a fundamental hydrodynamic phenomenon and serves as a benchmark problem in the analysis of hydrodynamic instabilities since the famous work of G.I. Taylor in 1923[5]. Recently, extensive studies have been focused on various variations, including axial TC flows, Taylor vortices in the transport phenomena in magnetic fluids, spherical Couette flows, viscoelastic TC flows, and so on[15-18]. Here we focus on the classical case where incompressible Newtonian fluids fill the gap between two concentric rotating infinitely-long cylinders. The outer cylinder is fixed while the inner one rotates (see Fig. 1(b) for device setup). When the fluid velocity is larger than some upper critical velocity, the system will transit from laminar flows to vortical flows and then eventually to turbulence. Conversely, it will stay stable as laminar flows when the fluid velocity is less than the lower critical velocity. When the velocity goes between the lower and upper critical velocities, it may be either laminar or turbulent.

Based on linear stability analysis of Navier-Stokes equations in cylindrical coordinates, a critical Reynolds number $Re = \Omega r_i d/\nu$ characterizing the stability of TC flows can be determined (see SI for details), where $d = r_o - r_i$ denotes the gap width with $r_o$ and $r_i$ being the radius of outer and inner cylinders respectively, $\Omega$ is the angular velocity of the inner cylinder, and $\nu$ is the kinetic viscosity. It is far more convenient to determine whether the flow is laminar or turbulent by identifying the critical points $Re^c$. When $Re < Re^c$, the TC flow is stable; when $Re$ is increased above but very close to $Re^c$, the flow instability caused by the curved streamlines of the base flow produces axisymmetric Taylor vortices. Note that $Re^c$ is a function of the radius ratio $\eta = r_i/r_o$.

The Rayleigh-Bénard convection, ubiquitous in nature phenomena, is another extensively studied hydrodynamic model. It happens between two large parallel plates where the lower one is heated (see Fig. 1(e)). The change in density due to temperature variations gives rise to a flow generated by buoyancy. This motion is opposed by viscous forces in the fluid and the force balance determines whether the flow is stable or not. If the temperature gradient, and thus the density gradient, is large enough, the gravitational force will dominate and the instability will occur. According to the linear stability analysis, when the Rayleigh number $R_a = \alpha(T_1 - T_0)gh^3 / (\upsilon\kappa)$ is larger than some critical value, the stability of RB flows will be lost (see SI). Here $\alpha$, $g$, $\upsilon$ and $\kappa$ denote the thermal expansion coefficient, gravitation constant, kinematic viscosity, and thermal diffusivity separately, $T_0$ and $T_1$ are temperatures of lower and upper plates, and $h$ is the distance between two plates.

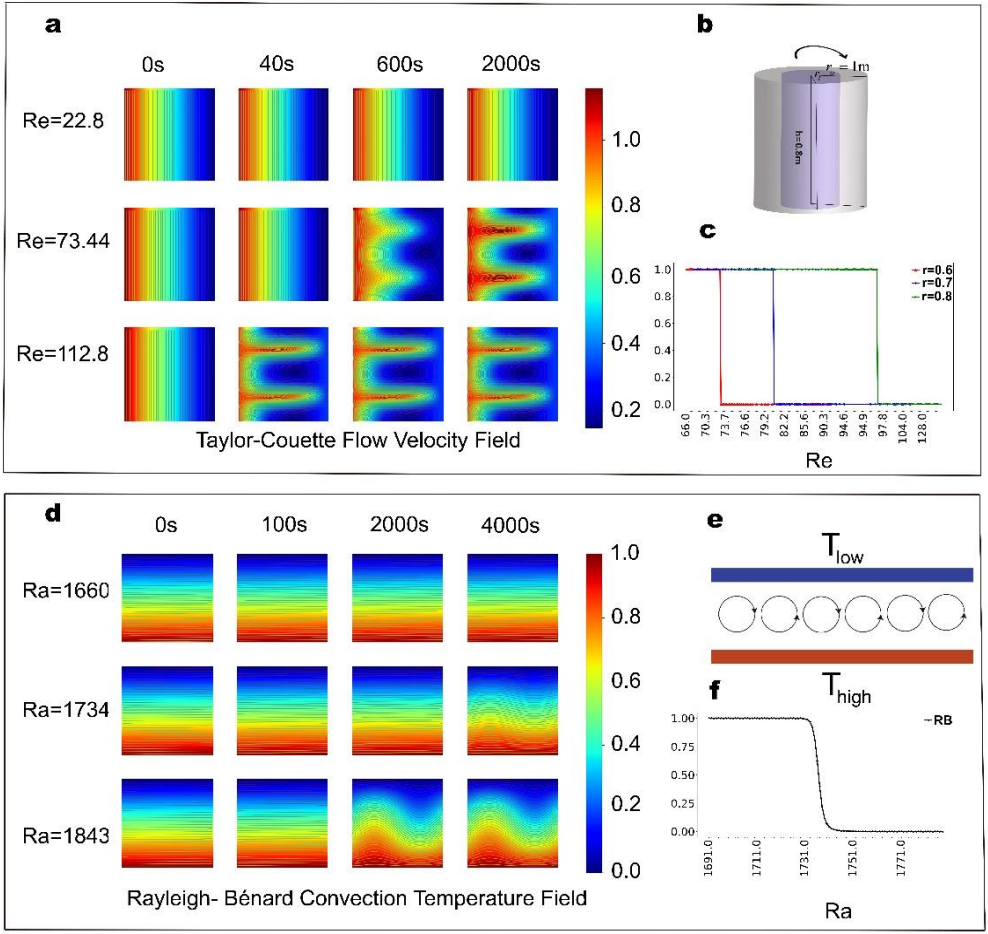

**FIG. 1**. (a)(d) Typical velocity and temperature fields before, during and after the transition point for TC flow and RB convection respectively. (b)(e) Sketches of experiment setup. (c)(f) Curves of transition point prediction based on CNN. For TC flows, three different radius ratios from 0.6 to 0.8 are used.

In the current work, typical TC flows and RB flows under various *Re* and *Ra* numbers are numerically simulated by an open-source CFD software package — OpenFOAM, which provides standard numerical solvers to hydrodynamic problems[19]. For TC flows, three different radius ratios from 0.6 to 0.8 are simulated with the Reynolds number ranging from around 2 to more than 100. For RB flows, the fluid flows are simulated for Rayleigh numbers from 1500 to 1900. For each Reynolds/Rayleigh number, we let the simulation run enough long time to make the flow to reach stationary. Based on OpenFOAM simulations, TC flows keep to be laminar when *Re* is less than 73 ( for $\eta$=0.6). Above the critical *Re* number, Taylor vortices are generated and the velocity field becomes dramatically different from that in laminar flows. For RB flows, the critical *Ra* number is around 1734, which provides a discrimination of the laminar flows and vortical flows (see Fig. 1).

This observation is directly confirmed by the K-means algorithm (see Fig. A1 and A2 in SI) and Mahalanobis distance, both of which show that the flow fields are divided into three clusters – laminar flows, near transition, and vortical flows, according to the Reynolds/Rayleigh number. Based on the Mahalanobis distance, the correlation between data before and after transition of TC flow (or RB convection) is relatively low, while the correlation among data before transition point (or after transition point) is obviously much higher (see Fig. 2).

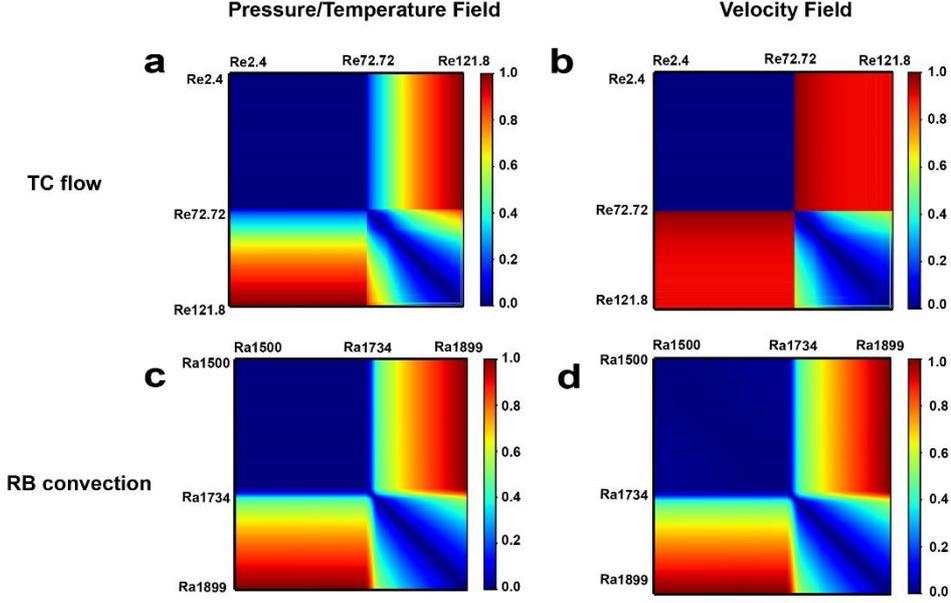

**FIG. 2**. Correlation analysis for (a,b)TC flow and (c,d)RB convection based on Mahalanobis distance.

### B. Hydrodynamic instability identified by CNN

To detect the transition from laminar flow to vortical flow in TC and RB flows, we refer to the convolution neural network (CNN)[20], which has a great merit to deal with data with complex spatial patterns, like image classification[21]. Since our classification is also based on the flow fields, *e.g.* the velocity, pressure and temperature distributions in space and time, we believe CNN is a suitable machine learning framework for this problem. To avoid taking advantage of information about transition point in advance, the training set are carefully prepared and only data with *Re/Ra* numbers far away from the critical points are taken (see Table A1 in SI). The classification results are depicted in Fig. 1(c) and 1(f), with transition points summarized in Table I.

For TC flows, the critical Reynolds numbers predicted by CNN are very close to those by LSA. Furthermore, the predicted transition curves as shown in Fig. 1(c) are so sharp that means CNN indeed has the capability to accurately distinguish laminar flows from vortical flows. The same conclusion also holds for the RB flows. This time the predicted critical *Ra* number by CNN is a slightly smaller than that LSA. However, this does not necessarily mean CNN gives a poor prediction since nonlinearity may either stabilize or destabilize a hydrodynamic flow, which may make the prediction of LSA deviate from the true values.

**TABLE I.** Critical transition point predicted by linear stability analysis and CNN.

| Flow type | Radius ratio | LSA | CNN |
| --- | --- | --- | --- |
| TC flow | η=0.8 | Re=96.0 | 96.0 |
|  | η=0.7 | Re=79.5 | 80.6 |
|  | η=0.6 | Re=71.7 | 72.7 |
| RB convection |  | Ra=1740.0 | 1734.0 |

### C. Robustness and noise-tolerance

The artificial neural network is a mimic of real biological neural network[22], especially human visual nervous system which is good at induction, analogy and promotion and shows great robustness and noise tolerance. For example, people can quickly recognize old friends even they have not seen each other for years. Therefore, we also hope CNN can inherit those merits. To this end, two experiments -- data masking by random placed filters and data mixed with noise, are carried out.

Firstly, we perform the data masking procedure. We randomly select 10% ~40% positions in a given flow field, and set the corresponding velocity and pressure/temperature values to be zero. From Fig. 3(a) and 3(c), it is clear seen that low masking ratio (less than 25%) shows little influence on the classification, while high masking ratio (greater than 40%) makes the classification on data above transition point less trustable (since we set masked values as zero). As a result, we can safely conclude that CNN has a moderate tolerance on data masking, which confirms the fact that CNN uses global features of the flow field rather than some local features to make judgment and classification, since the local information is more easily destroyed or lost during data masking.

Secondly, for a primitive flow field $X_{n\times p}$, random noises $N_{n\times p}=\alpha\|X_{n\times p}\|_{\infty}\eta$ are added to test the robustness and noise-tolerance of CNN, where $\alpha$=0.01-0.1 denotes the noise level, $\|X_{n\times p}\|_{\infty}$ is the maximal norm of $X_{n\times p}$, $\eta \sim \mathrm{N}(0,1)$ is a random number subject to the normal distribution. Again, CNN shows a quite extraordinary tolerance against random perturbations of the input data. And correct predictions are made even in the presence of 10% noise (see Fig. 3(b) and 3(d)), under which little useful information can be extracted by human's eyes.

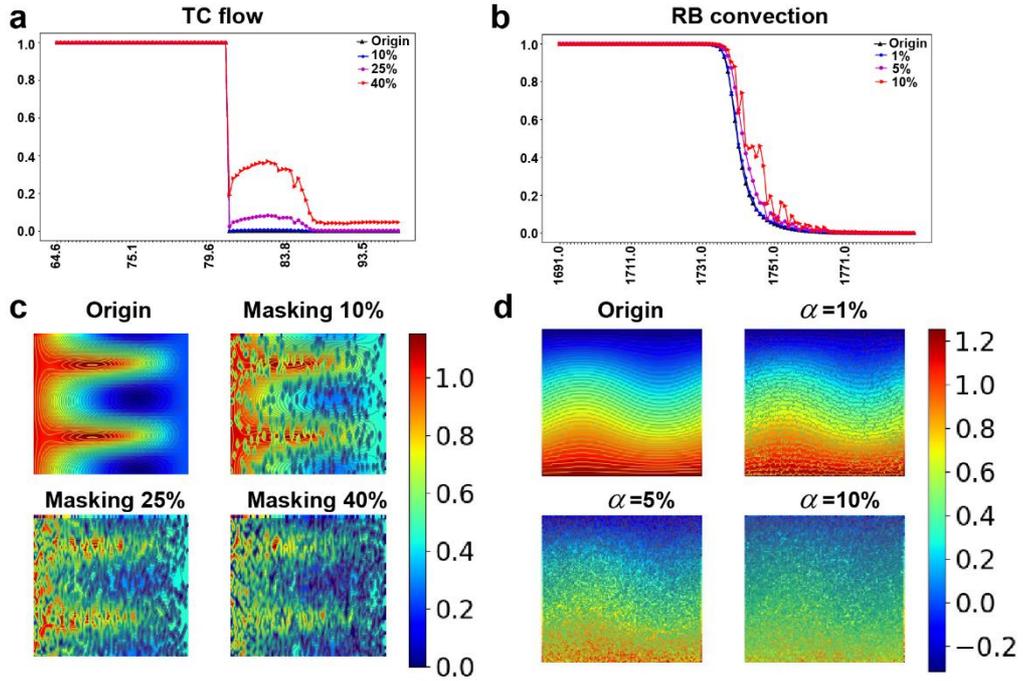

**FIG. 3**. Robustness tests of CNN against data masking and noise for (a) TC and (b)RB flows respectively. (c) Typical velocity fields for TC flow with masking ratio from 10% to 40%, and (d) temperature fields for RB flow as the noise level increases from 1% to 10%.

### D. Key spatial features revealed by PCA

Considering recent astonishing achievements of neural networks in diverse fields, we are not surprised to see the excellent performance of CNN in the detection of hydrodynamic instabilities. However, it is still not clear why CNN can work so well, which is also a big mystery in other applications. To provide a better understanding, we use Principal

Component Analysis (PCA)[23] to extract the key spatial features of the flow fields which provide an efficient low-dimensional representation of the raw data and will be used by CNN to determine the critical transition point as we believed.

In PCA, the eigenvalue represents the variance of the dataset on its corresponding eigenvector. The larger the eigenvalue is, the more dispersed the dataset will be along the direction of the eigenvector. So the eigenvector for the largest eigenvalue points out the direction in which the dataset makes most dramatic changes. In other words, it is also the most informative. During the determination of the critical transition point, we think CNN also takes advantage of this information, since the key issue is how to discriminate two different flow motions in nature, like the laminar and vortical flows, which is reflected by spatial variations in the velocity, pressure and temperature fields.

As shown in Fig. 4(b), the largest eigenvalue in PCA makes a contribution of 99.83% for TC flows and 92.14% for RB convection, with the corresponding eigenvector highlighting the major spatial differences between laminar and vortical flows. Due to the cyclic boundary conditions of TC flows in $z$ direction and RB flows in x direction, the eigenvectors also show a mirror symmetry.

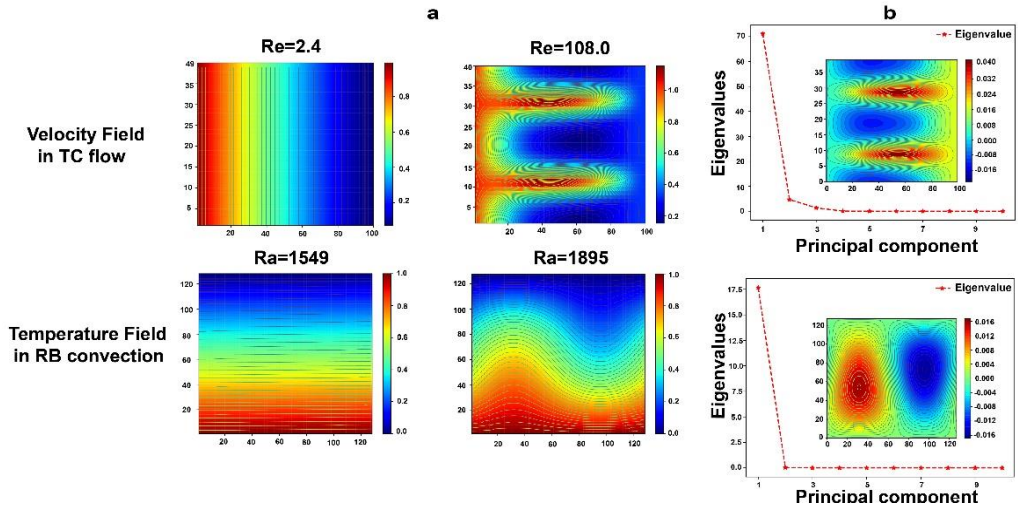

**FIG. 4**. (a) Typical velocity fields for TC flow and temperature fields for RB convection before and after the transition point. (b) Eigenvalue distribution for TC and RB flows by PCA respectively. In each case, the eigenvector corresponding to the largest eigenvalue is drawn in the subplot, which almost perfectly highlights the difference in flow fields before and after the transition point.

**TABLE II.** CNN for TC and RB flows.

| | |
|---:|:---|
| **Goal:** | Identify the onset of hydrodynamic instabilities and predict the critical transition point. |
| **Data:** | Typical TC and RB flows are simulated by OpenFoam. Stationary solutions of TC flows ($\eta$=0.6) with Re ranging from 10.8-65.9 and 91.3-120.0 (1562 data points) and of RB flows with Ra ranging from 1500-1650 and 1750-1899 (3289 points) are taken into the training set, while stationary solutions of TC flows with Re from 66.0 to 91.2 (68 points) and of RB flows with Ra from 1651 to 1749 (99 points) are taken as the test. |
| **Setup:** | The input data (velocity norm plus pressure/temperature) are normalized between 0 and 1. The convolutional neural network is composed by two convolution layers (kernel size of filters as 5*5) and two max pooling layers (kernel size as 2*2). After a fully connected layer, a probability |

value between 0 to 1 is exported as an indication of hydrodynamic instability.

| | |
|---|---|
| **Results:** | The laminar flows and vortical flows are well separated from each other by CNN, with the predicted transition points close to results by linear stability analysis. Reliable predictions are made even in the presence of 25% data masking ratio or 10% random noise. |
| **Significance:** | Identify the onset of hydrodynamic instabilities of TC and RB flows by neural network for the first time. |

## III. METHODS

### A. Setup of numerical simulations

#### 1. Taylor-Couette flow

The device setup of TC flow is sketched in Fig. 1(b). The height and radius of outer cylinder are fixed with $h = 0.8\,m$ and $r_o = 1.0\,m$, while the radius of inner cylinder is varied as $r_i = 0.6, 0.7, 0.8\,m$, resulting in a radius ratio $\eta = 0.6\text{-}0.8$. Thanks to the symmetry of TC flow along azimuth $\theta$ direction in cylindrical coordinate $(r, \theta, z)$, the computational domain of 3D flows between two concentric rotatory cylinders is simplified to a fan-shaped sector. Mathematically, it means the annulus $\{(r, \theta, z) \mid r_i \leq r \leq r_o, 0 \leq \theta < 360°, 0 \leq z \leq h\}$ simplifies to a wedge $\{(r, \theta, z) \mid r_i \leq r \leq r_o, 0 \leq \theta \leq 5°, 0 \leq z \leq h\}$, which largely reduces the computational cost. Correspondingly, the boundary patches (not the boundary conditions) in OpenFOAM are specified as *wedge* of front and back in θ-azimuth direction, *cyclic* of top and bottom in z-axial direction, and *patch* of inner-wall and outer-wall in the r-radical direction. By $N$, $K$ and $M$ we denote the respective numbers of mesh points along the radius, z-axis and azimuth. For numerical convenience, the whole computational region is uniformly divided into $100 \times 40 \times 1$ cells along the r-radius, z-axis and azimuth respectively.

Towards the boundary and initial conditions, periodic boundary conditions (BCs) in z-axial direction are adopted to reduce computational cost. That is, the cylinder top and bottom are equipped with periodic BCs to simulate an infinite long axis. The *wedge* BCs in azimuth $\theta$ direction are used due to its intrinsic symmetry to construct the computational domain of a whole cylinder from a wedge shape. No-slip BCs for velocity fields are imposed at the cylinder inner and outer surfaces (see Table A3 in SI). As to the initial condition, the exact steady solution corresponding to a fixed rotating angular velocity of the inner wall is used (see Eq. A4 in SI).

In OpenFOAM, the icoFoam solver for the incompressible, isothermal flow is used, with the time step $\Delta t = 0.001 s$ and kinematic viscosity $\nu = 0.01 m^2 s^{-1}$. Though the pisoFoam solver is more preferred for turbulent flows when the Reynolds number is much larger than the critical transition value, it demands a heavy computational cost and is not considered in the current work.

#### 2. Rayleigh–Bénard convection

The basic simulation setup for RB convection is shown in Fig. 1(e). The cuboid consists of a rectangular bottom of length $l = 4.03\,m$ and breadth $b = 0.01\,m$ in the $xz$-plane and the height $h = 0.1\,m$ in y-axis. Correspondingly, the boundary patches for RB convection are specified as *wall* of top and bottom in y-axis, *cyclic* of left and right in x-axis, and *empty* of front and back in z-axis. The *empty* patch requires no solution in its normal direction and thus leads to a pseudo-3D problem. For RB convection, the whole computational domain is uniformly divided into cells with $(N, K, M) = (256, 64, 1)$. By $N, K, M$, we denote the respective numbers of mesh points along xyz- Cartesian coordinates.

Towards the boundary and initial conditions, we take the temperature field as an illustration. The top temperature is maintained at a fixed value, while the bottom temperature is varied. Periodic BCs are imposed at the left and right surfaces. The front and back surfaces are equipped with *empty* BCs for a pseudo-3D case. The initial conditions for all variables are uniformly set to be zero.

To simulate the RB convection, we adopt the transient solver -- buoyantBoussinesqPimpleFoamsolver for buoyant, turbulent flows of incompressible fluids. The kinematic viscosity is set to be $v=0.01 m^2 s^{-1}$, the thermal expansion coefficient is $0.01$ $K^{-1}$, and the time step is $\Delta t = 0.025 s$.

### B. Convolutional Neural Network

The input velocity and pressure/temperature fields for TC and RB flows can be viewed as 2D pictures. Therefore, CNN is a natural choice for this problem due to its outstanding performance in image processing and no need for complex operations such as image preprocessing and feature extraction as in other neural network frameworks. So far, there are many classical convolutional neural network designs reported in literature, such as LeNet[24] which is used for digit recognition firstly, AlexNet[25], VGGNet[26], ResNet[27] and so on. The basic architecture of CNN we used is shown in Fig. 5, with details described below.

The CNN takes a normalized two-dimensional array (velocity plus pressure/temperature fields) as input, which is followed by a series of combined convolution and pooling operations. The kernel size of each convolutional filter is $5 \times 5$, with a stride of size 1 when the filter slides over the convolutional layer. The channel number of the convolutional layer is increased from 16 to 32, which enables the network to gradually learn more abundant features of flow fields. A max-pooling layer of size $2 \times 2$ follows the convolutional layer. In addition to introducing local translation invariance into the output, max-pooling also reduces the amount of computations. In the end, the output is connected to a fully connected layer.

In order to make the output $V$ a probability value, we process each output using the softmax function $\hat{y}_i = e^{V_i} / \sum_{i=0}^{K} e^{V_i}$, where $V_i$ is the output of the fully connected layer. $i$ is the category index, and the total number of categories is $K$. $\hat{y}_i$ describes the ratio of i'th element to the sum of all elements. During training, the cross entropy $L = -\sum_{i=1}^{K} y^{(i)} \log(\hat{y}^{(i)})$ is taken as the loss function, where $y^{(i)}$ is one-hot code (only one bit is allowed to be 1, other bits must be 0) of the true classified[28]. Adjustment of hyper-parameters is performed by Adam[29], with the learning rate set as 0.001. The whole CNN framework is implemented by PyTorch[30], with all codes available at https://github.com/yangwuyue/Fluid-phase-transitions/.

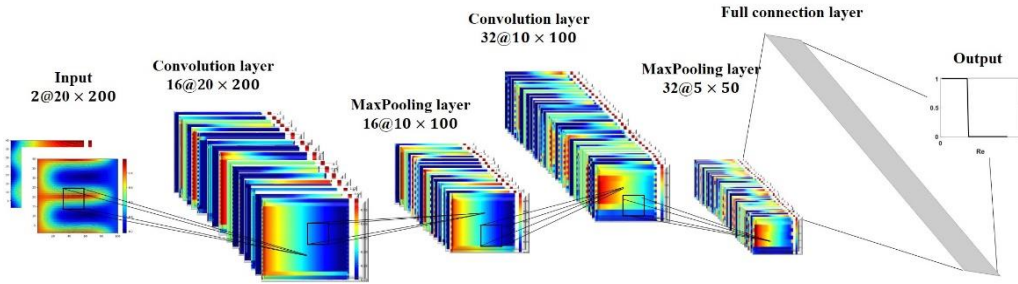

**FIG. 5**. Architecture of the convolutional neural network used in this study.

### C. Feature analysis

#### 1. Principal Component Analysis

We implement the principle component analysis by using the covariance method. Given the flow field $X = \left(x_{ij}\right)_{n \times m}$ where $n$ is the sample size and $m$ is the field dimension, we calculate the eigenvalues and eigenvectors of covariance matrix $\Sigma_{ij} = \frac{1}{n-1}\left(X_{ik} - \mu_k\right)\left(X_{kj} - \mu_j\right)$, where $\mu_j = \frac{1}{n}\sum_{i=1}^{n} X_{ij}$. Denote its eigenvalue and corresponding eigenvector as $\lambda_i$ and $v_i$ respectively, then $\Sigma v_i = \lambda_i v_i$. The eigenvalues are sorted from large to small as $\lambda_1 \geq \lambda_2 \geq \cdots \geq \lambda_p \geq 0$, and $\lambda_i / \sum_{n=1}^{p} \lambda_n$ stands for the variance contribution.

#### 2. Correlation Analysis

The correlation of dataset from laminar and vortical flows, can be obtained by calculating their Mahalanobis distance. For a dataset $X$ with mean value $\mu$ and covariance matrix $\Sigma$ as defined in PCA, the Mahalanobis distance of $x$ is defined as $D_M(X) = \sqrt{(X-\mu)^T \Sigma^{-1} (X-\mu)}$.

## IV. CONCLUSION

As a bridge connecting different types of flow motions, *e.g.,* laminar, vortical and turbulent flows, hydrodynamic instability is an important subject in modern fluid mechanics. Classically, the analysis of hydrodynamic instability is carried out by analytical approaches, like linear or nonlinear stability analysis, while in this work we propose an automatic solution based on pattern recognition handled by machine learning algorithms.

By inputting the flow fields, like velocity, pressure and temperature, the critical transition points for TC flow and RB convection are predicted by convolution neural network. The sharp sigmodal-like transition curves and well agreements with results based on linear stability analysis confirm the applicability of CNN in this problem. Furthermore, by principle component analysis, key spatial features charactering the variations in the flow fields before and after transition are revealed, which are then used by CNN to discriminate different types of fluid motions and to identify the critical transition points. This provides a deep understanding on how CNN can make reliable predictions. We believe our results will inspire more insightful studies and applications of machine learning algorithms in the field of fluid mechanics, especially the problems related to hydrodynamic instabilities, phase transition and phase separation.

## ACKNOWLEDGEMENTS

This work was supported by the National Natural Science Foundation of China (Grant Nos. 21877070 and 11871299) and the 13th 5-Year Basic Research Program of CNPC (2018A-3306). We thank for Prof. Wen-An Yong, Dr. Zhiting Ma and Dr. Juntao Huang for their helpful discussions. We thank Dr. Pipi Hu for his kind assistance with OpenFOAM.

**Declaration of Interests.**

The authors declare no conflicts of interest.